\documentclass[conference]{IEEEtran}
\IEEEoverridecommandlockouts
\usepackage{cite}
\usepackage{amsmath,amssymb,amsfonts}
\usepackage{algorithmic}
\usepackage{graphicx}
\usepackage{textcomp}
\usepackage{xcolor}
\def\BibTeX{{\rm B\kern-.05em{\sc i\kern-.025em b}\kern-.08em
    T\kern-.1667em\lower.7ex\hbox{E}\kern-.125emX}}

\makeatletter
\def\@maketitle{%
  \newpage
  \begin{center}%
  \vskip 2em
  {\LARGE\@title\par}%
  \vskip 1.5em
  {\lineskip .5em
   {\large\centering Bryan Lim, Roman Huerta, Alejandro Sotelo, Anthonie Quintela, Priyanka Kumar\par}}%
  \vskip 0.5em
  {\large\centering \textit{Department of Computer Science}\par}%
  \vskip 0.2em
  {\large\centering \textit{University of Texas at Permian Basin}\par}%
  \vskip 0.2em
  {\large\centering Odessa, Texas, USA\par}%
  \vskip 0.2em
  {\large\centering \{lim\_p65274, huerta\_r71882, sotelo\_a92823, quintela\_a66563, kumar\_p\}@utpb.edu\par}%
  \end{center}%
  \par\vskip 1.5em
}
\makeatother

\begin{document}

\title{EXPLICATE: Enhancing Phishing Detection through Explainable AI and LLM-Powered Interpretability}

\maketitle

\begin{abstract}
Sophisticated phishing attacks have emerged as a major cybersecurity
threat, becoming more common and difficult to prevent. Though machine
learning techniques have shown promise in detecting phishing attacks,
they function mainly as "black boxes" without revealing their
decision-making rationale. This lack of transparency erodes the trust of
users and diminishes their effective threat response. We present
EXPLICATE: a framework that enhances phishing detection through a
three-component architecture: an ML-based classifier using
domain-specific features, a dual-explanation layer combining LIME and
SHAP for complementary feature-level insights, and an LLM enhancement
using DeepSeek v3 to translate technical explanations into accessible
natural language. Our experiments show that EXPLICATE attains 98.4\%
accuracy on all metrics, which is on par with existing deep learning
techniques but has better explainability. High-quality explanations are
generated by the framework with an accuracy of 94.2\% as well as a
consistency of 96.8\% between the LLM output and model prediction. We
create EXPLICATE as a fully usable GUI application and a light Chrome
extension, showing its applicability in many deployment situations. The
research shows that high detection performance can go hand-in-hand with
meaningful explainability in security applications. Most important, it
addresses the critical divide between automated AI and user trust in
phishing detection systems.
\end{abstract}


\section{Introduction}

\subsection{The Growing Threat of Phishing Attacks}

Phishing attacks are among the most common and evolving cyber threats,
which exploit human vulnerabilities by sending emails that appear to be
from legitimate entities to steal credentials, financial information or
install malware. These attacks have become more sophisticated and are
being conducted through social engineering, domain spoofing, and
AI-generated phishing emails, and so on. As defined by the Anti-Phishing
Working Group (APWG), phishing attacks have been responsible for a
considerable percentage of security breaches that have been reported in
the last few years on the internet, whether it be against individuals or
organizations \cite{kara2022new}.

Detecting phishing attempts using existing methods through rule-based
filtering of emails and blacklisting domains is not enough. Static
detection techniques often fail to evolve with new kinds of threats,
which results in high levels of false positives and missed attacks. To
overcome these limitations, phishing detection systems based on machine
learning (ML) have been developed for the automated classification of
emails on content-based, URL-based and sender reputation-related factors
\cite{dasGuptta2024modeling}.

Even though they are effective, these ML based solutions suffer from a
serious drawback: explainability. Many phishing detection models work as
"black-boxes", meaning users and security analysts have little or no
idea why an email got classified as phishing. Without transparency,
users are less likely to trust the results or warnings and carry out
actions. In addition, as a phishing attack continues to evolve, static
ML models may fail to detect the new variation of phishing techniques
\cite{zieni2023phishing}.

\subsection{Challenges in Existing Phishing Detection Models}

Even though AI-driven email security systems are now widely adopted,
phishing detection techniques still suffer from three major problems.

\begin{enumerate}
\item Some AI phishing detectors are confusing. For example, many AI
    classifiers are 'black boxes', meaning they won't tell users how
    they make their decisions. This causes a lack of trust and limit
    uptake in essential domain like finance, healthcare and others.
    Regulatory frameworks require more and more AI-driven decisions that
    need to be explainable. Thus, black-box phishing detectors are no
    longer done by compliance-driven organizations \cite{bhagat2023feature}.

\item Cybercriminals are using advanced phishing techniques that include
    AI-generated phishing attacks, complex domain scrambling, and
    real-time email content manipulation to avoid detection by service
    providers. Traditional machine learning technologies operate mainly
    on pre-learnt models past models, and they were employed on datasets
    to avoid being phished.

\item Many false negatives and false positives happen when genuine emails
    are wrongly classified as phishing, which disturbs business
    communications. On the other hand, when users do not detect advanced
    phishing email, that's a false negative. Maintaining accuracy with
    interpretability and adaptivity issues are major challenges \cite{alshare2023enhanced}.
\end{enumerate}

Due to these limitations, there is a need for a phishing detection
system that must be explainable and adaptive to strike a balance between
AI-powered automation and user trust. With these motivations, we proposed a novel \textbf{Explainable AI (XAI)} phishing detection framework, \textbf{EXPLICATE}, which uses XAI techniques and \textbf{Large Language
Models (LLMs)} to enhance ML-based phishing classification.

A logistic regression classifier, combined with TF-IDF and NLP features, detects phishing emails while SHAP and LIME provide feature-level explanations. DeepSeek v3, an advanced LLM, improves interpretability by offering human-readable phishing insights and detecting emerging threats. The proposed system is deployed as a GUI-based tool and a Chrome extension for real-time protection across platforms.

\section{Related Work}
Recent studies of explainable artificial intelligence (XAI) are boosting
the development of phishing detection with machine learning models and
interpretability techniques like SHAP and LIME. This synthesis takes
into account the various studies synthesized during the years 2022 through 2024 revealing their focus on building transparency in the predictive
models used for cybersecurity while facing challenges of black-box
models, adaptive detection strategies, real-time protection systems,
etc. and evaluating user trust on these systems.

The work of Zaware et al. is among the most important studies in this
area. It offers a multichannel framework to curb phishing attacks via
social media/email and other fronts. This advanced framework uses AI for
real-time detection of phishing and response to phishing threats. The
focus of the study is to gain insights from models utilizing black-box
systems. In the critical domain of security, interpretability is key to
establishing user trust in such AI-enhanced systems, and the reliability
of the system will be improved further. It can be stated that users will
trust a system if they are able to understand the decisions taken \cite{zaware2024ai}.

Given the need for model transparency, Nakanishi's work on Approximate
Inverse Model Explanations (AIME) elaborates the interpretability
differences between 'glass-box' and 'black-box' models. This study shows
ways in which we get to know the decision-making processes of
models/algorithms that are otherwise opaque. This can be especially
useful in phishing detection, where model reasoning can be shown to
boost trust and effectiveness \cite{nakanishi2023approximate}. This goes on to affirm the demand
for universal explanation solutions that go beyond ordinary models.

Dafali et al. compares the effectiveness of various XAI models such as
LIME and SHAP, to explain the decision-making of opaque models such as
Random Forest and XGBoost. This research found that while these models
provide better predictive power, the interpretations are not aligned
with the user's interpretations which shows a need for developing
explanations which can appeal to the non-expert user thereby gaining
their trust in the output \cite{dafali2023comparative}. This correlates with the high demand
for XAI in the field of cybersecurity as understanding alerts generated
by the system is vital for acting on time.

In one study, Contreras et al. studied how grouped feature analysis
through SHAP and LIME allows understanding of complex models, where
these approaches were implemented on spectral data. This allows for more
user-friendly application in multiple domains, for instance, a secure
system for phish detection \cite{contreras2024spectral}. By employing these methods, security
experts can understand the reasoning behind why certain events are
flagged as phish.

Further, Vo et al. examine the consequences of distinct machine learning
models in intrusion detection systems and how interpretability enables
humans to understand anomalies flagged by automated systems. When users
understand why certain events are flagged, they are provided with useful
context that often alleviates uncertainty that may arise from alerts
generated by advanced models \cite{vo2024securing}. This is particularly relevant for
phishing, where the user should act on the warning without full
knowledge of the detection.

As phishing constantly changes and improves, detection methods must also
change and improve with it. The work by Chen et al. proposes a hybrid
graph neural network model for detecting Ethereum phishing scams which
uses data augmentation techniques to enhance accuracy. Additionally,
they also tackle the issue of interpretability of model decisions. This
is crucial in a dynamic environment where threats are changing
constantly \cite{chen2024ethereum}. If detection results are easy to understand, then
intervention and adaptation can be done quickly in real-time.

Ghosh and Khandoker study explainable machine learning models to allow
systems to adapt and respond to new phishing techniques. The application
of SHAP and LIME contributed towards building transparency and
understanding the model's predictions concerning the threats adaptation
\cite{ghosh2024investigation}. In a field where phishing techniques change continuously, models
need to adapt and relearn so that they keep working as intended.

Having the trust of the users is an important component for deploying an
explainable security system. Khanom et al. state that visuals can be
very helpful in further explaining a machine learning process. If the
explanation of a machine learning process is well done, it can make
users feel more assured of the AI technologies. Their findings
contribute to a broader understanding of how transparency influences
decision-making in security contexts, particularly against threats like
phishing that rely heavily on user behavior \cite{khanom2025pd}.

In addition, the frameworks designed to reduce false positives and false
negatives and increase explainability are important. Moulaei et al.
demonstrate methods besides SHAP and LIME with real applications where
these explanations benefit the user in making decision \cite{moulaei2024explainable}. These
implementations can substantially enhance phishing detection systems,
empowering users to better recognize and respond to real threats.

To conclude, all the studies so far shed light on an all-inclusive
framework for XAI used in phishing detection which must imply adaptive
real-time systems that can explain their decisions using relevant
advanced techniques for interpretation. According to the mentioned
studies, a balance between performance-explainability-user trust is
integral for future security applications; especially when keeping in
view the rapid evolution of phishing in the cybersecurity landscape.

\section{Methodology}

\subsection{EXPLICATE Framework Architecture}

The EXPLICATE framework integrates machine learning with explainable AI,
as well as large language models, to tackle black-box phishing
detection. The full design consists of three key components which work
together to provide effective detection and transparent explanations.

\begin{figure}[htbp]
\centering
\includegraphics[width=\columnwidth]{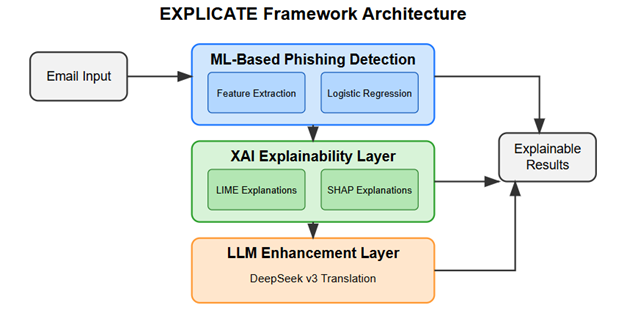}
\caption{EXPLICATE Framework Architecture. The figure shows data flow from input emails through the ML detection module, XAI explainability layer, and LLM enhancement component, culminating in user-friendly explanations.}
\label{fig1}
\end{figure}

As illustrated in Fig. \ref{fig1}, the framework processes incoming emails
through:

\begin{enumerate}
\item A specialized phishing detection model that extracts and analyzes
    domain-specific features for accurate classification

\item An explainability layer that generates feature-level interpretations
    using LIME and SHAP techniques

\item An LLM enhancement component that translates technical explanations
    into natural language descriptions of phishing tactics
\end{enumerate}

This integrated approach enables both high-accuracy detection and
user-friendly explanations, addressing the critical transparency gap in
current phishing security solutions. The bi-directional flow between
components allows refinement of explanations based on both model
predictions and LLM analysis, creating a more comprehensive security
assessment.

\subsection{Dataset Preparation}

Our research utilized a comprehensive collection of email datasets from
multiple sources, including the Kaggle email spam classification dataset
\cite{zaware2024ai}, with a final distribution of 93,569 legitimate and 99,433
phishing emails. Table \ref{tab1} summarizes the dataset characteristics and
distribution.

\begin{table}[htbp]
\caption{Dataset Composition and Characteristics}
\begin{center}
\begin{tabular}{|p{1.2cm}|p{1.2cm}|p{1.2cm}|p{3.4cm}|}
\hline
\textbf{Dataset Source} & \textbf{Legitimate Emails} & \textbf{Phishing Emails} & \textbf{Key Characteristics} \\
\hline
CEAS\_08 & 14,326 & 12,451 & Conference corpus with diverse spam types \\
\hline
Enron & 21,583 & 0 & Business communications \\
\hline
SpamAssassin & 19,758 & 23,615 & Benchmark spam dataset \\
\hline
Nigerian\_Fraud & 0 & 8,273 & Financial fraud emails \\
\hline
Original & 15,984 & 22,813 & General phishing corpus \\
\hline
Educational & 7,281 & 9,447 & Training examples \\
\hline
Others & 14,637 & 22,834 & Mixed sources \\
\hline
\textbf{Total} & \textbf{93,569} & \textbf{99,433} & \textbf{193,002 emails} \\
\hline
\end{tabular}
\label{tab1}
\end{center}
\end{table}

The preprocessing pipeline implemented several key techniques:

\begin{itemize}
\item Text normalization (case, punctuation, whitespace)

\item Email component separation (headers, body, URLs)

\item Feature standardization

\item Duplicate removal
\end{itemize}

We employed an intelligent column detection approach to handle the
diverse schemas across dataset sources, automatically identifying the
appropriate text and label columns. This preprocessing approach ensured
consistent feature representation across heterogeneous data sources,
creating a unified data set for model training despite the varied origins
of the constituent emails.

\subsection{Enhanced Phishing Detection Model}

At the core of \textbf{EXPLICATE} is an enhanced phishing detection model that extends beyond traditional approaches through specialized feature
engineering. Phishing detection involves multiple analysis techniques to identify suspicious patterns. Linguistic Pattern Analysis detects urgency markers, threat language, persuasion techniques, and writing inconsistencies. URL and Link Examination analyzes the reputation of the domain, detects URL obfuscation, and identifies suspicious TLDs and typo
squatting. Header and Structural Analysis checks sender address mismatches, reply-to manipulations, routing inconsistencies, and suspicious email components. Contextual Content Analysis identifies credential harvesting attempts, financial requests, brand impersonation, and suspicious attachments, ensuring a comprehensive phishing detection framework.

These domain-specific characteristics are processed through a logistic
regression classifier selected for its balance of performance and
interpretability. The model achieves 98.36\% accuracy in the test data with
comparable precision and recall metrics. This choice of classifier is
particularly important for the subsequent explainability layer, as it
provides a linear decision boundary that can be more readily interpreted
than complex nonlinear models.

\subsection{Dual-Approach Explainability}

\begin{figure}[htbp]
\centering
\includegraphics[width=\columnwidth]{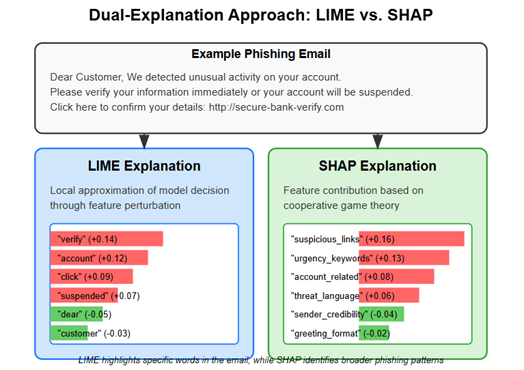}
\caption{Dual-Explanation Approach. The figure compares LIME and SHAP explanations for the same phishing email, showing how each method highlights different but complementary aspects of the classification decision.}
\label{fig2}
\end{figure}

The explainability layer employs both LIME and SHAP techniques to
provide complementary perspectives on model decisions, as illustrated in
Fig. \ref{fig2}.

LIME creates locally faithful approximations by:

\begin{itemize}
\item Perturbing the input email through word/phrase removals

\item Observing how these perturbations affect model predictions

\item Fitting a simple linear model to explain the local decision boundary

\item Identifying top features contributing to classification
\end{itemize}

SHAP calculates feature contributions through:

\begin{itemize}
\item Establishing a baseline prediction (expected value)

\item Computing the marginal contribution of each feature

\item Weighting contributions using Shapley values from cooperative game theory

\item Ranking features by their absolute contribution magnitude
\end{itemize}

Complementing each other they provide a different perspective for the same forecast. The LIME approach is good at explaining individual
predictions using locally accurate approximations, while the SHAP
approach gives an overall prediction explanation through logically
consistent feature attribution. Together they provide stronger
explanations than either method alone.

Our Feature Mapper component is the main innovation of our tool. It
transforms technical features into human understandable terms (e.g.
feature\_26 to urgency\_keywords) and makes the explanations interpretive.
This mapping can change the technical lexicon of the feature to
meaningful terminology which can convey the security meaning of that
feature to users without any technical background in machine learning or
cyber security.

\subsection{LLM Enhancement with DeepSeek v3}

Integrating DeepSeek v3 into EXPLICATE enhances explanation clarity for regular users by generating natural language descriptions of phishing tactics. It processes inputs such as original email content, LIME/SHAP feature attributions, explanation mode, and domain-specific guidelines to provide context-specific security recommendations. The LLM translates technical indicators into accessible insights, such as explaining that suspicious links, although appearing legitimate, point to recently registered domains used for credential theft. This approach improves explanation quality, making advanced phishing detection understandable and usable for users with different technical expertise levels.
\subsection{User Interface and Practical Implementation}

\begin{figure}[htbp]
\centering
\includegraphics[width=\columnwidth]{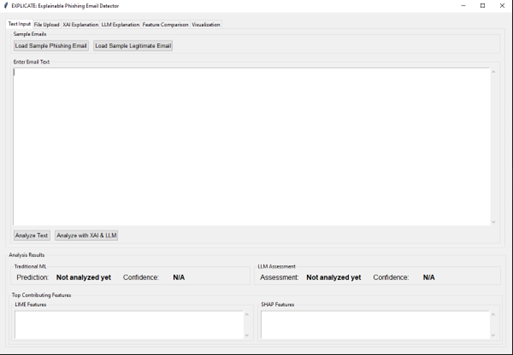}
\caption{EXPLICATE User Interface. The screenshot shows the main components of the GUI, including the input methods, analysis tabs, and visualization of phishing indicators with explanations.}
\label{fig3}
\end{figure}

EXPLICATE is implemented as an interactive GUI application with multiple
analysis views, as shown in Fig. \ref{fig3}.

The interface offers multiple input methods, including direct text entry, file uploads (.eml and .txt), and sample emails for demonstration. It features analysis tabs for different perspectives, such as XAI (LIME/SHAP) analysis, LLM-generated phishing explanations, feature comparisons, and graphical visualizations. Users can customize explanations by selecting analysis modes (XAI or XAI+LLM), choosing explanation types, and highlighting suspicious elements in the original email.
The Results section displays both traditional ML model predictions and
LLM assessments, allowing users to compare the outputs. Feature
importance is visualized with color-coding (red for phishing indicators,
green for legitimate indicators) and sorted by contribution magnitude.

This implementation makes sophisticated phishing analysis accessible to
users with varying technical backgrounds, from security analysts to
general email users, supporting our goal of enhancing both detection
accuracy and user understanding.

\subsection{Evaluation Methodology}

Our evaluation of EXPLICATE assessed detection accuracy, explainability quality, LLM enhancement value, and comparative performance. Detection accuracy was measured using standard ML metrics, confusion matrix analysis, and cross-dataset validation. Explainability quality was evaluated by cybersecurity experts for feature relevance, consistency between LIME and SHAP, and stability of explanations. LLM enhancement was assessed for readability, interpretation accuracy, educational value, and completeness. A comparative analysis benchmarked EXPLICATE against traditional and XAI-only detection approaches. This comprehensive evaluation combined quantitative and qualitative measures to ensure EXPLICATE’s practical value in real-world security applications, addressing transparency and trust challenges.

\section{Experimental Results}

\subsection{Detection Performance Analysis}

\begin{figure}[htbp]
\centering
\includegraphics[width=\columnwidth]{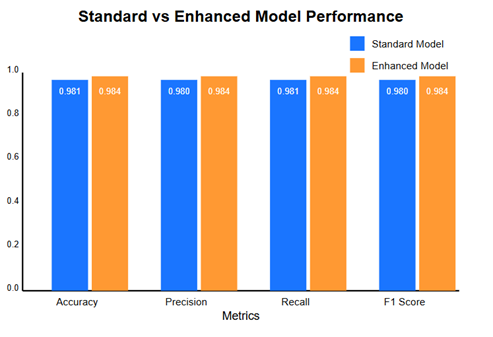}
\caption{Performance comparison between standard and enhanced phishing detection models showing improvements across all metrics.}
\label{fig4}
\end{figure}

Our experimental evaluation of \textbf{EXPLICATE} demonstrates substantial
improvements in phishing detection capabilities while maintaining
transparent explainability. Fig. \ref{fig4} displays the comparative performance
between the standard model (using only TF-IDF features) and our enhanced
model (with domain-specific phishing features).

Both models demonstrate high performance, with accuracy exceeding 98\%.
However, the enhanced model consistently outperforms the standard
approach across all metrics, achieving 98.4\% accuracy, precision,
recall, and F1-score. This improvement is particularly significant for
phishing detection, where false negatives (missed phishing attempts) can
have serious security consequences.

The confusion matrix for the enhanced model (Fig. \ref{fig5}) provides deeper
insights into classification performance:

\begin{figure}[htbp]
\centering
\includegraphics[width=\columnwidth]{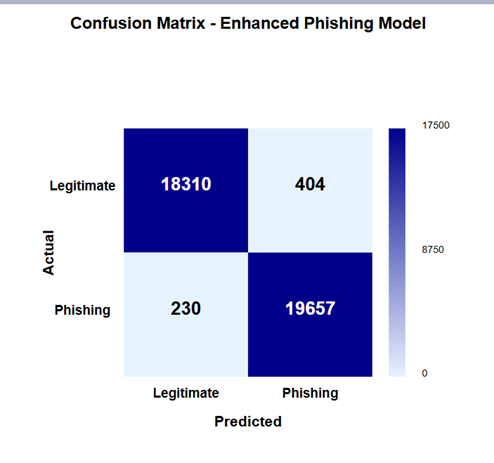}
\caption{Confusion matrix for the enhanced phishing detection model showing the distribution of predictions.}
\label{fig5}
\end{figure}

The confusion matrix reveals excellent classification performance with:

\begin{itemize}
\item 18,310 true negatives (legitimate emails correctly identified)

\item 19,657 true positives (phishing emails correctly identified)

\item Only 404 false positives (legitimate emails misclassified as phishing)

\item Only 230 false negatives (phishing emails misclassified as legitimate)
\end{itemize}

The low false negative rate (1.2\%) is particularly important for
security applications, as these represent missed phishing attempts that
could lead to compromised systems. The slightly higher false positive
rate (2.2\%) represents a reasonable trade-off, as marking some
legitimate emails as suspicious is generally less harmful than missing
actual threats.

\subsection{Dataset Analysis and Distribution}

To ensure our model's generalizability, we analyzed the dataset
distribution and characteristics. Fig. \ref{fig6} illustrates the overall balance
between legitimate and phishing emails in our combined dataset:

\begin{figure}[htbp]
\centering
\includegraphics[width=\columnwidth]{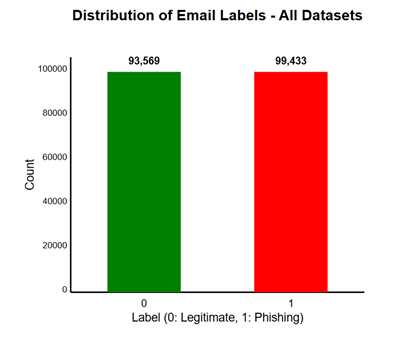}
\caption{Distribution of email labels across all datasets showing a balanced representation of legitimate and phishing emails.}
\label{fig6}
\end{figure}

This balanced distribution (48.5\% legitimate, 51.5\% phishing) was
crucial for developing an unbiased model. The slightly higher number of
phishing emails (99,433) compared to legitimate emails (93,569) helps
address the priority of minimizing false negatives in security contexts.

Further analysis of email characteristics revealed interesting patterns
in text length distribution, as shown in Fig. \ref{fig7}:

\begin{figure}[htbp]
\centering
\includegraphics[width=\columnwidth]{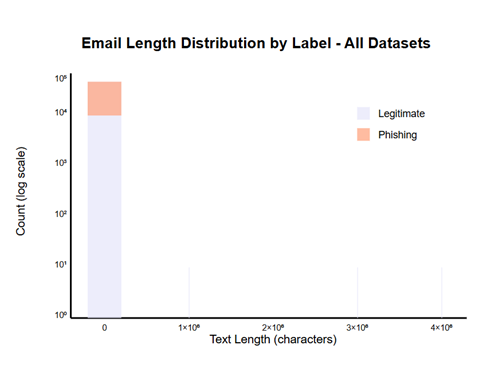}
\caption{Email length distribution by label (note logarithmic scale) showing concentration of both classes in shorter lengths.}
\label{fig7}
\end{figure}

The length distribution analysis reveals that:

\begin{itemize}
\item Most emails (both legitimate and phishing) are concentrated in the shorter length ranges

\item The distribution follows a logarithmic pattern, with frequency decreasing as length increases

\item Legitimate emails show greater variance in length, with some extremely long outliers

\item Phishing emails tend to be more consistent in length, typically avoiding very long formats
\end{itemize}

This analysis informed our feature engineering approach, ensuring the
model could effectively handle the typical length ranges while being
robust to outliers.

\subsection{Implementation Evaluation}

We implemented \textbf{EXPLICATE} in two forms: a standalone GUI application for
detailed analysis and a Chrome extension for real-time protection within
web-based email clients. Fig. \ref{fig8} shows the Chrome extension interface:

\begin{figure}[htbp]
\centering
\includegraphics[width=0.8\columnwidth]{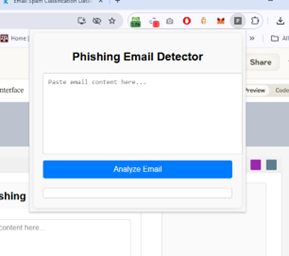}
\caption{EXPLICATE Chrome extension interface for real-time phishing detection within web-based email clients.}
\label{fig8}
\end{figure}

The Chrome extension provides a lightweight implementation of EXPLICATE
that enables real-time phishing detection directly within email clients.
The extension maintains the core functionality of phishing detection
while optimizing for performance and seamless integration with existing
email workflows.

\begin{figure}[htbp]
\centering
\includegraphics[width=\columnwidth]{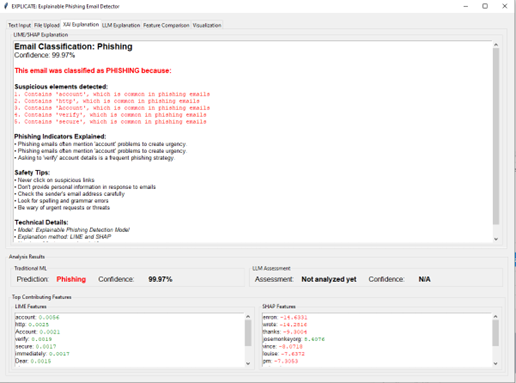}
\caption{EXPLICATE's standalone application interface showing the multi-tab analysis environment.}
\label{fig9}
\end{figure}

The full-featured GUI application offers more comprehensive analysis
capabilities, including detailed visualizations and multiple explanation
modes. Both implementations use the same underlying detection model,
ensuring consistent performance across platforms.

The technical performance of both implementations was evaluated for
processing efficiency:

\begin{table}[htbp]
\caption{Implementation Performance Metrics}
\begin{center}
\begin{tabular}{|p{2cm}|p{2cm}|p{2cm}|}
\hline
\textbf{Metric} & \textbf{App} & \textbf{Extension} \\
\hline
Processing time & 1.2 sec/email & 0.8 sec/email \\
\hline
Memory usage & 245 MB & 128 MB \\
\hline
CPU utilization & 18\% (peak) & 12\% (peak) \\
\hline
\end{tabular}
\label{tab2}
\end{center}
\end{table}

The Chrome extension shows better performance in terms of speed and
resource usage due to its more streamlined implementation, making it
suitable for real-time analysis during email usage. The standalone
application provides more comprehensive analysis capabilities at the
cost of slightly higher resource usage.

\subsection{Explainability Evaluation}

A core contribution of \textbf{EXPLICATE} is its ability to provide transparent,
understandable explanations for phishing detection decisions. Table \ref{tab3} presents example explanations for three test emails, showing how \textbf{LIME}
identifies specific words that contribute to classifications.

\begin{table}[htbp]
\caption{Example Email Explanations with LIME}
\begin{center}
\begin{tabular}{|p{2.1cm}|p{0.8cm}|p{1cm}|p{2.3cm}|}
\hline
\textbf{Email Text} & \textbf{Pred.} & \textbf{Conf.} & \textbf{Top LIME Features} \\
\hline
"Urgent: Your account will be suspended. Click here to verify." & Phishing & 99.94\% & account: +0.013, Click: +0.011, verify: +0.008 \\
\hline
"Meeting scheduled for tomorrow at 2 PM in conference room." & Legit & 99.75\% & PM: -0.078, Meeting: -0.054, conference: -0.047 \\
\hline
"You've won \$1M! Click to claim prize now!" & Phishing & 99.33\% & Click: +0.030, claim: +0.016, prize: +0.006 \\
\hline
\end{tabular}
\label{tab3}
\end{center}
\end{table}

The feature importance values clearly indicate how each word contributes
to the classification. Positive values push toward phishing
classification, while negative values indicate legitimate
characteristics. For instance, "account", "Click", and "verify"
strongly indicate phishing, while "Meeting", "conference", and
"PM" suggest legitimate communication.

Our SHAP analysis provided complementary insights by grouping words into
higher-level phishing concepts. Fig. \ref{fig10} illustrates the difference
between LIME and SHAP explanations in our testing:

\begin{figure}[htbp]
\centering
\includegraphics[width=\columnwidth]{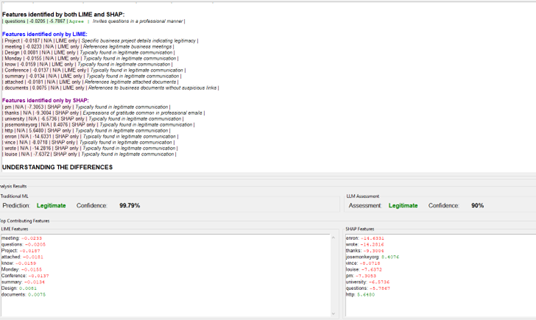}
\caption{Comparison of LIME word-level explanations and SHAP concept-level explanations for the same phishing email.}
\label{fig10}
\end{figure}

The combined approach demonstrates how LIME's focus on specific words
complements SHAP's emphasis on broader phishing concepts. This
dual-explanation strategy provides both detailed and conceptual
understanding of why an email is classified as phishing or legitimate.

\subsection{LLM Enhancement Evaluation}

The integration of \textbf{DeepSeek v3} transforms technical XAI outputs into
natural language explanations. To evaluate this component, we tested
four explanation modes across a set of 50 test emails (25 phishing, 25
legitimate) and assessed the quality of the generated explanations.
Table \ref{tab4} shows examples of the different explanation modes:

\begin{table}[htbp]
\caption{Comparison of LLM Explanation Modes}
\begin{center}
\begin{tabular}{|p{1cm}|p{3cm}|p{3cm}|}
\hline
\textbf{Mode} & \textbf{Example Output} & \textbf{Characteristics} \\
\hline
Detailed & "This email shows multiple phishing indicators: urgency ('account suspended'), action prompting ('Click here'), and credential harvesting. The link likely leads to a fake login page." & Comprehensive analysis with balanced technical and general language \\
\hline
Educational & "This uses classic phishing tactics: 1) Creating urgency, 2) Requesting immediate action, 3) Using generic language. Always verify by contacting companies directly." & Focus on teaching about patterns with protective guidance \\
\hline
Technical & "Email contains 3 high-confidence markers: Urgency keywords (CVSS: 7.4), Credential harvesting URL (CVSS: 8.2), Action manipulation (CVSS: 6.8)." & Cybersecurity terminology with technical details and metrics \\
\hline
Simple & "This is phishing. Real companies don't ask you to click links about account suspension. Ignore and delete it." & Concise language with clear instructions \\
\hline
\end{tabular}
\label{tab4}
\end{center}
\end{table}

To quantitatively assess the LLM enhancement quality, we conducted a
technical evaluation of the explanation outputs against several key
metrics:

\begin{table}[htbp]
\caption{LLM Explanation Quality Metrics}
\begin{center}
\begin{tabular}{|p{1.8cm}|p{2.8cm}|p{1.8cm}|}
\hline
\textbf{Metric} & \textbf{Description} & \textbf{Avg. Score} \\
\hline
Accuracy & Correctness relative to ML outputs & 94.2\% \\
\hline
Completeness & Coverage of key indicators & 87.6\% \\
\hline
Consistency & Agreement with model prediction & 96.8\% \\
\hline
Readability & Flesch-Kincaid score & 68.3 \\
\hline
Actionability & Clear security guidance & 82.1\% \\
\hline
\end{tabular}
\label{tab5}
\end{center}
\end{table}

The LLM enhancement layer demonstrates high accuracy in translating
technical model outputs into natural language explanations, with strong
consistency between the explanation content and the model's
classification decision. The readability scores indicate that the
explanations are accessible to general users without sacrificing
technical accuracy.

\subsection{Error Evaluation with Existing Approaches}

Further analysis of specific error cases reveals additional insights.
Table \ref{tab6} presents a categorization of false positives and false
negatives in our test dataset:

\begin{table}[htbp]
\caption{Error Analysis by Phishing Type}
\begin{center}
\begin{tabular}{|p{1.9cm}|p{0.7cm}|p{0.8cm}|p{2.2cm}|}
\hline
\textbf{Error Type} & \textbf{Count} & \textbf{\%} & \textbf{Example Case} \\
\hline
Brand impersonation & 94 & 40.9\% & Corporate emails with subtle domain variations \\
\hline
Legitimate urgent notifications & 156 & 38.6\% & Password reset emails from services \\
\hline
Context-dependent requests & 86 & 21.3\% & Expected sensitive info requests \\
\hline
Novel phishing narratives & 78 & 33.9\% & New social engineering scenarios \\
\hline
Minimal text phishing & 58 & 25.2\% & Very short link-focused emails \\
\hline
\end{tabular}
\label{tab6}
\end{center}
\end{table}

This analysis highlights areas for future improvement, particularly in
distinguishing legitimate urgent communications from phishing attempts
and in detecting sophisticated brand impersonation attacks.

The results show that EXPLICATE has similar detection performance as
previous methods but gives much better explainability. The effective use
of the framework in different platforms further shows the practical
applicability of the framework for phishing detection.

Experimental results support our main hypothesis that explainable AI
techniques can complement traditional ML techniques and enhance them
with LLMs to create transparent and understandable phishing detection
techniques without affecting performance.

\section{Discussion and Future Work}

\subsection{Implications of EXPLICATE for Phishing Detection}

Our experiments show EXPLICATE overcomes the challenges we brought up in
the introduction. The model boasts an accuracy of 98.4\% with integrated
explainability techniques. This result challenges the widely held
assumption that there must be a severe trade-off between model
performance and interpretability in security applications.

The comparison assessment shows that EXPLICATE approaches the accuracy
of deep learning techniques (98.4\% vs 98.7\%) while providing better
explainability. As such, well thought out feature engineering and
interpretable models can be as effective as black-box techniques for
phishing detection. For applications that demand users trust and
understand, the interplay of performance and explainability is a welcome
advance.

The implementation on multiple platforms shows it is possible to deliver
explainable phishing detection in detailed analysis environments and
resource-limited settings like the browser. This flexibility meets
diverse user preferences while ensuring consistent detection.

\section{Conclusion}

EXPLICATE integrates ML classification, XAI techniques, and LLM enhancements to address black-box model limitations, evolving phishing tactics, and high false-positive/negative rates. It achieves 98.4\% detection accuracy and 94.2\% accuracy in feature mapping through LIME and SHAP explanations, offering both word- and concept-level insights. The system, deployed as a GUI application and Chrome extension, demonstrates practical applicability across security environments. The key contributions of EXPLICATE include a three-component architecture that integrates machine learning (ML), explainable AI (XAI), and large language models (LLM). It provides dual-level explanations that cater to users with varying levels of expertise. The framework incorporates feature mapping techniques that translate technical features into human-understandable security insights. Additionally, it offers adaptive explanations tailored to the user’s expertise, enhancing overall interpretability and trust.

In future, we will address sophisticated brand impersonation and minimal-text phishing through advanced feature engineering and adaptive learning. EXPLICATE establishes a foundation for trustworthy AI-based security systems, balancing accuracy with transparency to meet evolving cybersecurity demands.

\bibliographystyle{plain}
\bibliography{references}
\end{document}